\documentclass[prd,longbibliography,nofootinbib,twocolumn,preprintnumbers]{revtex4-1}

\usepackage[T1]{fontenc}
\usepackage[utf8]{inputenc}
\usepackage{relsize}
\usepackage{soul}
\usepackage{calligra}
\usepackage{verbatim}
\usepackage{caption}
\usepackage{comment}
\usepackage{microtype}
\usepackage[american]{babel}
\usepackage{booktabs}

\usepackage{adjustbox}

\usepackage[titles]{tocloft}
\cftsetindents{section}{0em}{2.5em}
\cftsetindents{subsection}{2.5em}{2.5em}

\usepackage{appendix}

\usepackage{graphicx,epsfig}
\usepackage[usenames,dvipsnames,table]{xcolor}
\usepackage{subfig}
\usepackage{float}
\usepackage{placeins}

\usepackage{tabularx}
\usepackage{makecell,multirow}
\usepackage{dcolumn}
\usepackage{enumitem}

\usepackage{amsmath,amsthm,amssymb}
\usepackage{slashed}
\usepackage{mathrsfs}
\usepackage{bm}
\usepackage{leftidx}
\usepackage{commath}

\usepackage{url}
\usepackage{color}
\definecolor{darkblue}{rgb}{0.0,0.0,0.6}
\definecolor{red}{rgb}{0.9, 0,0}
\definecolor{navy}{rgb}{0.05, 0.05,0.8}
\definecolor{linkcolor}{rgb}{0.0, 0.28, 0.67}

\usepackage[colorlinks]{hyperref}
\hypersetup{
     colorlinks   = true,
     citecolor    = linkcolor,
     linkcolor    = linkcolor,
     urlcolor    = linkcolor
}
\newcommand{\Eq}[1]{Eq.~(\ref{eq:#1})} 
\newcommand{\Eqs}[2]{Eqs.~(\ref{eq:#1}) and~(\ref{eq:#2})} 
\newcommand{\Sec}[1]{Sec.~\ref{sec:#1}} 
\newcommand{\Secs}[2]{Secs.~\ref{sec:#1} and \ref{sec:#2}} 
 
\newcommand{\Fig}[1]{Fig.~\ref{fig:#1}}

\newcommand{\x}{\chi}

\newcommand{\eV}{\text{eV}}

\newcommand{\GeV}{\text{GeV}}
\newcommand{\TeV}{\text{TeV}}
\newcommand{\V}{\text{V}}
\newcommand{\kV}{\text{kV}}
\newcommand{\MV}{\text{MV}}

\newcommand{\cm}{\text{cm}}

\newcommand{\m}{\text{m}}
\newcommand{\km}{\text{km}}
\newcommand{\s}{\text{s}}

\newcommand{\Hz}{\text{Hz}}

\newcommand{\be}{\begin{equation}}
\newcommand{\ee}{\end{equation}}

\newcommand{\order}[1]{\mathcal{O}{(#1)}}

\newcommand{\grad}{\nabla}

\newcommand{\p}{\prime}

\newcommand{\Ap}{A^\prime}
\newcommand{\mAp}{m_{A^\prime}}
\newcommand{\eps}{\epsilon}

\newcommand{\subdm}{_{_\text{DM}}}
\newcommand{\mdm}{m\subdm}
\newcommand{\rhodm}{\rho\subdm}
\newcommand{\vdm}{v\subdm}
\newcommand{\w}{\omega}
\newcommand{\Zeff}{Z_\text{eff}}

\newcommand{\qx}{q_\x}
\newcommand{\mx}{m_\x}
\newcommand{\px}{p_\x}
\newcommand{\Ecr}{E_\text{cr}}
\newcommand{\Lx}{d}

\newcommand{\xv}{{\bf x}}

\preprint{FERMILAB-PUB-24-0002-SQMS-T, USTC-ICTS/PCFT-24-08}

\begin{document}

\preprint{FERMILAB-PUB-24-0002-SQMS-T}
\preprint{USTC-ICTS/PCFT-24-08}

\title{Millicharged Condensates on Earth}

\author{Asher Berlin$^{a,b}$}
\email{aberlin@fnal.gov}
\author{Roni Harnik$^{a,b}$}
\email{roni@fnal.gov }
\author{Ying-Ying Li$^{c,d}$}
\email{ yingyingli@ustc.edu.cn}
\author{Bin Xu$^{e}$}
\email{binxu@pku.edu.cn}
\affiliation{$^a$Theory Division, Fermi National Accelerator Laboratory}
\affiliation{$^b$Superconducting Quantum Materials and Systems Center (SQMS),
Fermi National Accelerator Laboratory}
\affiliation{$^c$Peng Huanwu Center for Fundamental Theory, Hefei, Anhui 230026, China}
\affiliation{$^d$Interdisciplinary Center for Theoretical Study, University of Science and Technology of China, Hefei, Anhui 230026, China}
\affiliation{$^e$School of Physics and State Key Laboratory of Nuclear Physics and Technology, Peking University, Beijing 100871, China}


\begin{abstract}

We demonstrate that long-ranged terrestrial electric fields can be used to exclude or discover ultralight bosonic particles with extremely small charge, beyond that probed by astrophysics. Bound condensates of scalar millicharged particles can be rapidly produced near electrostatic generators or in the atmosphere. If such particles directly couple to the photon, they quickly short out such electrical activity. Instead, for interactions mediated by a kinetically-mixed dark photon, the effects of this condensate are suppressed depending on the size of the kinetic mixing, but may still be directly detected with precision electromagnetic sensors. Analogous condensates can also develop in other theories involving new long-ranged forces, such as those coupled to baryon and lepton number.

\end{abstract}


\maketitle

\section{Introduction}
\label{sec:intro}

It has long been understood~\cite{Heisenberg:1936nmg,Sauter:1932gsa,Schwinger:1951nm} that strong electromagnetic fields drastically modify the structure of the vacuum. This gives rise to various nonlinear corrections to Maxwell's theory, which among their predictions include the spontaneous production of electron pairs in the presence of an electric field above the critical value $\Ecr \sim m_e^2 / e \sim 10^{12} \ \MV / \m$, the so-called Schwinger effect~\cite{Schwinger:1951nm}. Despite its strong theoretical footing, this process has never been observed, due to the difficulty in experimentally realizing such large fields.

In this work, we show that $\order{1}$ modifications to electromagnetic observables can occur at much lower field strengths if there exist bosonic millicharged particles (mCPs) $\x^\pm$ with mass $\mx$ and electric charge $\qx \ll 1$ much smaller than that of the electron. In this case, long-ranged electric potentials can be quickly shorted-out by the exponential Bose-enhanced growth of a bound millicharged condensate. As we show below, the growth of such a condensate is possible for $\qx \gtrsim (e E L^2)^{-1}$ and  occurs within a time $t \sim L \, \log{(1/\qx)}$, where $E$ is the strength of the electric field and $L$ is its coherence length. While discharging of sources via mCP pair-production was investigated previously in Refs.~\cite{Li:2013pfh,Hook:2017vyc,Korwar:2017dio}, the possibility of an enhanced bosonic population was not considered.

\begin{figure}
\centering
\includegraphics[width=0.48 \textwidth]{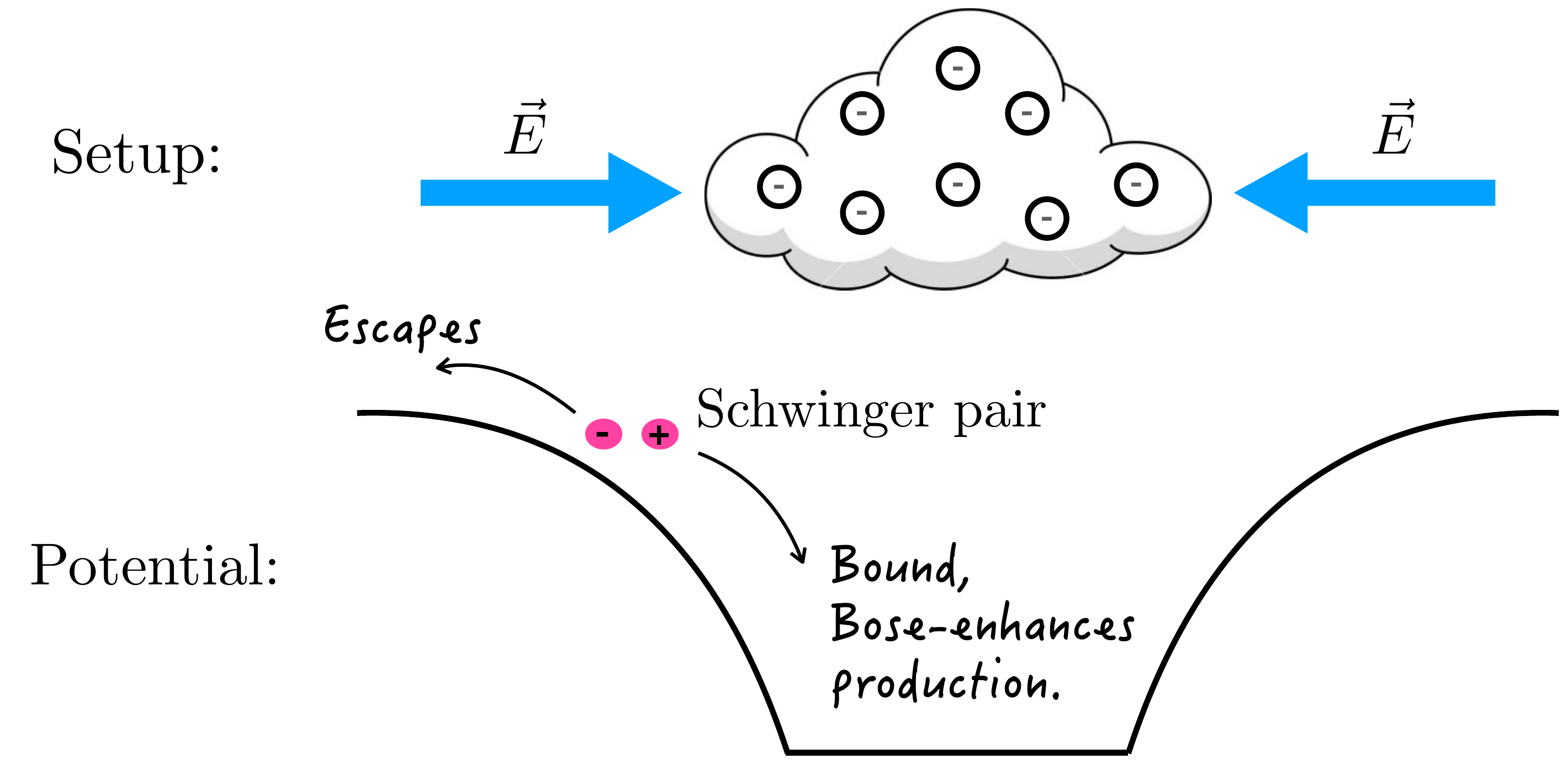}
\caption{A sketch of the setup (top) and the electrostatic potential (bottom). An object is electrostatically charged, generating a strong Coulomb field around it. Light scalar millicharged particles, if they exist as degrees of freedom, will be Schwinger pair-produced in the strong $E$-field. One sign of millicharges is expelled due to Coulomb repulsion, while the other remains electrostatically bound. The bound population will then Bose-enhance further pair-production, leading to a runaway behavior and the formation of a condensate, which ultimately screens the electric field.}
\label{fig:cartoon}
\end{figure}

The \emph{strongest} electric fields, as found in, e.g., superconducting cavities or ferroelectric materials, are either generated on short timescales or across small distances. Thus, although they can lead to mCP pair-production~\cite{Gies:2006hv,Berlin:2020pey}, they are not ideal for seeding the growth of Bose condensates. This is better suited for \emph{long-ranged} quasistatic electric fields, which arise from, e.g., atmospheric 
electricity~\cite{Feynman:1963uxa}. Indeed, Earth's fair-weather electric field $E_\oplus \sim 10^2 \ \V / \m$ spans a distance of $\sim 10^2 \ \km$ from the crust to the ionosphere. However, the presence of Earth's much larger magnetic field $B_\oplus \sim 10^2 \, E_\oplus$ typically hampers bosonic pair-production (we discuss an exception to this in \Sec{atmosphere}).  Thus, in this work, we focus predominantly on large Van de Graaff generators and on the electrical activity in thunderstorms, since these produce approximately quasistatic and long-ranged fields $E \sim 1 \ \MV / \m$ and $E \sim 10^2 \ \kV / \m$ much greater than $B_\oplus$ across distances of $L \sim 1 \ \m$ and $L \sim 1 \ \km$, respectively~\cite{PhysRev.43.149,Herb1959,ThunderBook}. In this case, such electric fields could be quickly discharged by the production of a bound condensate of scalar mCPs, as shown schematically in \Fig{cartoon}.

The remainder of this work is organized as follows. In \Sec{schwinger}, we discuss the production of an mCP Bose condensate. In \Sec{VanDeGraaff}, we investigate its consequences on large Van de Graaff generators, such as the one in use at the Boston Museum of Science, finding that its successful operation excludes the existence of ultralight bosonic mCPs that couple directly to the photon with charge $\qx \gtrsim 10^{-14}$. In \Sec{atmosphere}, we discuss how the generation of condensates in Earth's atmosphere is possible for much smaller couplings, $\qx \gtrsim 10^{-18}$. In \Sec{direct}, we discuss direct detection of millicharged condensates and show how modifications to our findings can arise due to model-dependent considerations, such as when mCP interactions are mediated by an ultralight kinetically-mixed dark photon. Finally, we conclude in \Sec{conclusion} and discuss directions for future work, such as analogous effects in theories incorporating new long-ranged forces coupled to baryon and lepton number. 

\section{Millicharged Bose Condensates}
\label{sec:schwinger}

In this section, we discuss the process of mCP Schwinger production. For more detailed reviews, see, e.g., Refs.~\cite{Kim:2000un,Kim:2003qp,Kim:2007pm,Hebenstreit:2011pm,Hebenstreit:2011pm,Hebenstreit:2011wk,Ruffini:2009hg,Kim:2019joy}. In most of this work, we treat the mCPs as genuinely charged under electromagnetism; in \Secs{atmosphere}{direct}, we discuss modifications when such interactions instead arise indirectly from a kinetically-mixed dark photon. Here, we also ignore additional processes which may dominate over the Schwinger effect, such as electromagnetic cascades that are triggered by energetic charged particles or photons (see, e.g., Ref.~\cite{Arvanitaki:2021qlj}). Our approach is thus conservative, as the effect of such additional reactions is to increase the mCP production rate. 

When the energy gained by a virtual particle-antiparticle dipole in an electric field exceeds that of its rest mass, the field is said to be supercritical and pair-production is possible. This is often discussed within the context of charged fermions. In the presence of supercritical Coulomb fields, the bound state energy levels are embedded in the Dirac sea continuum of negative-energy states.  It is thus energetically favorable for a particle in the continuum to transition to such a bound state, leaving behind an antiparticle hole. At the particle-level, this corresponds to pair-production, in which particles of one charge escape away to infinity, while the species of opposite charge remain localized near the interaction region, screening the source of the potential. Such pair-production marks the transition to the new charged ground state of the theory, which is stabilized by the Pauli principle.

A similar story also applies to the production of charged scalars, which will be the focus of this work. However, in this case the neutral vacuum evolves towards a highly-occupied Bose condensate that is stabilized by Coulomb interactions and is of sufficient charge to screen the total electric field to be subcritical~\cite{Klein:1976xj,Greiner,Rafelski:1978}.  The initially supercritical neutral vacuum is therefore unstable against decay to the new subcritical charged vacuum. In this section, we show that if there exist ultralight millicharged bosons, the transition to this subcritical state typically occurs within a tiny fraction of a second, corresponding to the generation of a millicharged condensate that can short-out the largest terrestrial electric fields, thus leading to observable consequences. 

\subsection{Requirements for Production}
\label{sec:production}

The requirements for Schwinger pair-production can be derived heuristically from basic kinematic arguments. The vacuum is populated by virtual mCP pairs separated by distances of $\sim (\mx^2 + \px^2)^{-1/2}$, where $p_\x$ is the mCP momentum. Pair-production is exponentially suppressed unless a background electric field that is coherent over a distance $L$ can accelerate this virtual pair to be on-shell, i.e., $e \qx E \, \Lx \gtrsim (\mx^2 + \px^2)^{1/2}$, where we have defined the distance accessible to this virtual pair as
\be
\label{eq:Lx}
\Lx \sim \min \big[ L \, , \, (\mx^2 + \px^2)^{-1/2} \big]
~.
\ee
Thus, the weakest requirement on the electric field occurs for $\px \ll \mx$. From this we infer that the critical field for mCP  Schwinger production is
\be
\label{eq:Ecr}
\Ecr \sim \frac{\mx}{e \qx} \, \max \big( m_\x \, , \, L^{-1} \big)
~,
\ee
where pair-production is exponentially suppressed for $E \ll \Ecr$.
This agrees with detailed studies of the Schwinger effect in inhomogenous electric fields~\cite{Kim:2007pm,Hebenstreit:2011pm,Hebenstreit:2011pm,Hebenstreit:2011wk,Ruffini:2009hg,Kim:2019joy}.

Note that for $E \gg \Ecr$, the above kinematic constraints are satisfied even for $\px \gg \mx$, corresponding to production of relativistic mCPs. In this case, we can drop the mass in the expressions above. Also, demanding that the mCP de Broglie wavelength fits inside the coherent electric field region $\px  L \gtrsim 1$, we find that for supercritical fields mCPs are produced with a characteristic momentum in the range
\be
\label{eq:momrange}
L^{-1} \lesssim \px \lesssim \sqrt{e \qx E}
~.
\ee
Self-consistency imposes that the upper limit is greater than the lower limit in \Eq{momrange}, which is equivalent to $\qx \gtrsim (e E \, L^2)^{-1}$. This, along with $E \gtrsim \Ecr$, form the basic requirements for the formation of an mCP condensate, which can be expressed as the following lower bound on the mCP charge,
\be
\label{eq:qxLL2}
\qx \gtrsim (e E)^{-1} \, \max \big[ \mx^2 \, , \,  L^{-2} \big]
~.
\ee
Note that the kinematic arguments leading to \Eq{qxLL2} can be equivalently rephrased as demanding that $E \gtrsim \Ecr$ and that both the so-called ``nucleation length'' $\sim \mx / (e \qx \, E)$ of the virtual mCP pair and the de Broglie wavelength after accelerating through the electric field $\sim 1 / (e \qx \, E L)$ are smaller than the characteristic scale $L$ of spatial variations in the electric field (see, e.g., Ref.~\cite{Arvanitaki:2021qlj} and references therein). Also note that \Eq{qxLL2} is simply the requirement that the interaction is in the strong coupling regime. To see this, consider an electric field generated by charge $Q$, $E = e Q / (4 \pi L^2)$, such that \Eq{qxLL2} requires $\Zeff \, \alpha  \gtrsim 1$, where $\Zeff \equiv |Q \, \qx|$.

If a strong magnetic field $B \gtrsim E$ is also present,  the growth of the bosonic condensate is hindered~\cite{Kim:2003qp}. This can be derived by a simple modification to the kinematic arguments above. In this case, the energy of a spin-0 particle is increased by the lowest-lying Landau level associated with its cyclotron motion, $\omega_L \sim e \qx \, B / (\mx^2 + \px^2)^{1/2}$. Thus, pair-production also requires $e \qx E \, \Lx \gtrsim  \omega_L$, which imposes the additional constraint $E \gtrsim B$.

Thus, if both \Eq{qxLL2} and $E \gtrsim B$ are satisfied, the vacuum decays to mCP pairs  initially localized within a distance $L$ of the electric source. As discussed below in \Secs{evolution}{KG}, in this case the mCPs that possess an attractive interaction with the source remain electrically bound, seeding a  condensate that grows in density exponentially. 

\subsection{Time Evolution of Bound States}
\label{sec:evolution}

At the microscopic level, the growth of the charged condensate corresponds to particle pair-production, after which only a single sign of charge becomes electrically bound. To remain bound near the source of the quasistatic supercritical potential, we impose that $p_\x \lesssim e \qx \, E L$. Note that this is  in fact already guaranteed from \Eqs{momrange}{qxLL2}. Thus, such mCPs with momenta in the full kinematic range of \Eq{momrange} remain localized near the source. The species of opposite charge is instead accelerated away to far distances, setting up a millicharge asymmetry. 

As the phase space of the bound population continues to grow, it quickly exceeds that of the unbound mCPs and stimulates further pair-production at an exponential rate. Steady state is attained only once the stable ground state of the theory is reached, corresponding to when the charge of the bound population efficiently screens the electric source to be subcritical~\cite{Klein:1976xj,Greiner,Rafelski:1978}.

Here, we show that this final configuration can be attained on microscopic timescales. Let us begin by reviewing the Schwinger production rate of charged scalar particles for supercritical electric fields $E \gg \Ecr$. Approximating the momentum distribution as isotropic and flat up to a maximum momentum of $p_\x \lesssim \sqrt{e \qx \, E}$, the phase space density of a bound species evolves according to $df_\x / dt \simeq \Gamma_\x \, (1+f_\x )$~\cite{Nishida:2021qta}. The growth rate is given by
\begin{align}
\label{eq:gamma0}
&\Gamma_\x \simeq \frac{3}{4 \pi} \, \frac{\Delta V}{V} \, \sqrt{e \qx \, E} 
~,
\end{align}
where $\Delta V / V \leq 1$ is a geometric volume penalty to be discussed below \Eq{rho}.
Integrating $f_\x$ over momentum and taking $f_\x \ll 1$ and $\Delta V / V = 1$ yields the standard form for the Schwinger rate~\cite{Kim:2003qp,Cohen:2008wz}.  

For mCPs $\x$ that remain electrically bound to the source, the phase space occupancy accumulates over time, leading to exponential growth within a time $t \sim \Gamma_\x^{-1}$,
\be
\label{eq:fpsol}
f_\x \simeq e^{\Gamma_\x  t} - 1 
~.
\ee
Instead, the oppositely-charged mCPs (denoted as $\bar{\x}$) are unbound and quickly evacuated from the region. Thus, their phase space only grows within the time $\Delta t \simeq L$ it takes such particles to exit the region of the coherent electric field. The phase space of this population is therefore
\be
f_{\bar{\x}} \simeq  \Gamma_\x \, \int_{t_0}^t d t^\prime ~  \big(1 + f_\x (t^\p) \big)
~,
\ee
where $t_0 \equiv \max{(0 , t - L)}$. This can be evaluated using \Eq{fpsol}, 
\be
\label{eq:fpbarsol}
f_{\bar{\x}} \simeq e^{\Gamma_\x  t} - e^{\Gamma_\x t_0} 
~.
\ee
Since we are interested in the charge density, we take the difference of \Eqs{fpsol}{fpbarsol},
\be
\label{eq:fpm}
f_\pm \equiv f_\x - f_{\bar{\x}} \simeq e^{\Gamma_\x t_0} - 1
~.
\ee
Integrating \Eq{fpm} over momentum space and multiplying by the per-particle charge then yields the total charge density
\be
\label{eq:rho}
\rho_\pm \simeq \frac{e \qx \, (e \qx \, E)^{3/2}}{6 \pi^2} ~ \big( e^{\Gamma_\x  t_0} - 1 \big)
~.
\ee

Typically, the volume $\Delta V$ over which pair-production can efficiently stimulate further production is much smaller than the total  volume occupied by the bound mCPs. In this case, the $e$-fold rate is suppressed by the ratio $\Delta V / V$, as in \Eq{gamma0}. For instance,  bound mCPs produced in a supercritical Coulomb field are quickly accelerated to a momentum beyond the upper limit of \Eq{momrange} within a radial distance of $\ell \simeq 1 / \sqrt{e \qx E}$, suppressing the degree of phase space overlap. Hence, the only mCPs that can Bose-enhance further production of mCPs at a radius $r$ are those produced within a narrow radial shell of thickness $\sim 2 \ell$ centered at $r$. When this length scale is much smaller than the range $L$ of the electric field, $\Gamma_\x$ is thus suppressed by this small volume contributing to Bose-enhanced production $\Delta V / V \simeq 2 \ell / L \simeq 2 / \sqrt{e \qx E L^2}$. Using this in \Eq{gamma0}, we find that for charges satisfying \Eq{qxLL2} the growth rate is parametrically $\Gamma_\x \simeq 3 / (2 \pi L)$, independent of the charge $\qx$. This is to be expected as the lower bound on the growth rate; after an mCP that is electrically bound to a quasistatic source traverses a full orbit, it returns to its initial phase space point, thus stimulating further production of mCPs within this same phase space region.

\subsection{Klein-Gordon in a Coulomb Potential}
\label{sec:KG}

Here, we verify these heuristic arguments by solving for the time-evolution of the mCP Klein-Gordon field in an attractive Coulomb-like potential (for related calculations, see, e.g., the various examples presented in Ref.~\cite{Greiner:2000cwh}). This is done by solving for the eigenfrequency $\w$, whose positive imaginary value indicates an instability to the theory and the exponential growth of the charged scalar field. This is therefore qualitatively similar to the process of superradiance~\cite{Ternov:1978gq,Zouros:1979iw,Detweiler:1980uk,Arvanitaki:2009fg,Arvanitaki:2010sy}, in which a spinning (and possibly charged~\cite{Bekenstein:1973mi,Bosch:2016vcp}) black hole gives rise to rapid growth of a gravitationally bound bosonic cloud. However, in our case, an important difference arises in the fact that the trapping potential is set solely by electromagnetic interactions and thus does not decouple in the small mass limit.   

The Klein-Gordon equation for a millicharged scalar field $\phi_\x (\xv , t) = e^{-i \w t} \, \phi_\x(\xv)$ coupled to an electrostatic potential $V(\xv)$ is given by
\be
\label{eq:KG1}
\Big[ \grad^2 - \mx^2 + \big( \w - V(\xv) \big)^2  \Big] \, \phi_\x (\xv) = 0
~.
\ee
For a spherically-symmetric potential $V(\xv) = V(r)$, the field can be factorized into radial and angular components $\phi_\x(\xv) = ( R(r) / r) \, Y (\theta, \phi)$, where $Y(\theta,\phi)$ are the usual spherical harmonics. Using this in \Eq{KG1}, the radial wavefunction satisfies
\be
\label{eq:KGr1}
\bigg( \frac{d^2}{dr^2} - \frac{\ell (\ell + 1)}{r^2} + k(r)^2 \bigg) \, R(r) = 0
~,
\ee
where $\ell$ is the standard azimuthal quantum number and we defined $k(r)^2 \equiv ( \w - V(r) )^2 - \mx^2$.

As a representative toy model, we adopt an attractive potential energy of the form of a spherical shell of charge $Q$ and radius $L$. In particular, we take
\be
V(r) = \begin{cases} 
- \Zeff \, \alpha / r & (r \geq L)
\\
- \Zeff \, \alpha / L & (r \leq L) ~, 
\end{cases}
\ee
where $\alpha$ is the fine-structure constant and $\Zeff = |Q \, q_\x|$ is the product of the shell's and scalar field's charge (in units of the electric charge).

Outside of the shell ($r > L$), $V(r)$ is a standard Coulomb potential. In this case, $R_\text{out}(r)  = c_\text{out} \, W_{\lambda, \mu} (\beta r)$, where $c_\text{out}$ is a normalization constant, $\beta \equiv 2 (\mx^2 - \w^2)^{1/2}$, and $W_{\lambda, \mu}$ is a Whittaker function with the subscripts defined as~\cite{Greiner:2000cwh}
\be
\label{eq:Rout}
\lambda \equiv \frac{2 \Zeff \, \alpha \, \w}{\beta}
~~,~~
\mu^2 \equiv \Big( \ell + \frac{1}{2} \Big)^2 - (\Zeff \, \alpha)^2
~.
\ee

Inside the shell ($r < L$), $V(r)$ is instead a constant. Solving \Eq{KGr1} and neglecting solutions singular at the origin, we find $R_\text{in}(r) = c_\text{in} \, r \, j_\ell (k_\text{in} r)$, where $c_\text{in}$ is a normalization constant, $j_\ell$ is a spherical Bessel function, and $k_\text{in}^2 \equiv ( \w + \Zeff \, \alpha / L)^2 - \mx^2$.

The eigenfrequency $\w$ is determined by matching the wavefunction and its first derivative at the shell boundary, which amounts to numerically solving $R^{\, \p}_\text{out} (L) / R_\text{out} (L) = R^{\, \p}_\text{in} (L) / R_\text{in} (L)$. For $\Zeff \, \alpha \ll 1$ and  $\mx \ll 1/(\Zeff \, \alpha \, L)$, this yields $\text{Im}(\w) = 0$ and the standard atomic energy levels, $\text{Re}(\w) \simeq \mx - (\Zeff \, \alpha)^2 / 2 n^2$, where $n$ is a positive integer. Instead, for $\Zeff \, \alpha \gg 1$ and $\mx \ll 1/L$, $\w$ develops a positive imaginary component $\Gamma_\x \equiv 2 \, \text{Im} (\w) > 0$, corresponding to exponential growth. In particular, in this case we find that the growth rate for the $\ell = 0$ state can be roughly approximated by
\be
\Gamma_\x \simeq \frac{1}{2L} \, \Big( 1 - \frac{\pi n}{\Zeff \, \alpha} \Big)
\ee
for $\text{Re}(\w) \simeq (- \Zeff \, \alpha + \pi n / 2) / L < 0$, where $n \lesssim \Zeff \, \alpha / \pi$ is a positive integer. This qualitatively agrees with the discussion in \Sec{evolution}.

Thus, the field and corresponding charge density $\rho_\pm$ grow exponentially, set by the rate $\Gamma_\x \sim 1 / L$, as in \Eq{rho}. This growth saturates when  $\rho_\pm$ screens the source, such that the total field is subcritical, i.e., once $\rho_\pm (t) \sim E / L$ within a time $t \sim L \, \log{(1/ \qx)}$.\footnote{Equivalently, the growth saturates once the work done in accelerating the unbound mCPs drains the electric energy. To see this, note that  $\rho_\pm \sim E  / L$ corresponds to a total of $N_\x \sim E \, L^2 / \qx$ produced pairs. The work done in ejecting the unbound mCPs is therefore $N_\x \, \qx \, E \, L \sim E^2 \, L^3$, comparable to the total energy stored in the initial electric field configuration.} We expect this growth to hold even in the presence of additional mCP reactions, provided that they conserve charge (such as annihilations to photons or dark photons), since these do not alter the total charge density. Furthermore, we do not expect scattering between mCPs and the environment to significantly alter the growth of the condensate. This is because for the values of $\qx$ considered here, the time after which an mCP  is likely to upscatter off of terrestrial matter and become unbound is many orders of magnitude longer than the time $\Gamma_\x^{-1} \sim L$ to grow the condensate by an additional $e$-fold.\footnote{Note that this is unlike black hole superradiance.  In that case, the instability is only present for a comparatively narrow range of $\text{Re}(\w)$, and the corresponding $e$-fold timescale $\Gamma_\x^{-1}$ is parametrically longer than the light-crossing time~\cite{Arvanitaki:2010sy}. As a result, the development of the superradiant instability is more sensitive to non-linearities from gravitational and self-interaction effects.}

The detailed spatial structure of this late-time steady state charge configuration is dictated parametrically by the momenta of mCPs and is stabilized by Coulomb self-interactions. For explicit studies along these lines, see, e.g., Refs.~\cite{Klein:1976xj,Greiner,Rafelski:1978}. Here, we are mainly interested in the timescale of the transition to this final state. In the sections below, we consider several examples of quasistatic electric sources, which can be discharged by mCPs well within the time it takes the condensate to grow by $\sim 10^2$ $e$-folds. Thus, the successful operation of these sources can be used to exclude the existence of ultralight bosonic mCPs directly charged under electromagnetism. 

\section{Van de Graaff Generators}
\label{sec:VanDeGraaff}

From \Eq{qxLL2}, we see that devices that operate large, long-ranged, and approximately static electric potentials are ideal for enhancing the sensitivity. In this section, we focus on the largest air-insulated Van de Graaff generator, which is operated for public display at the Theater of Electricity at the Boston Museum of Science~\cite{bostonmuseum}. 

This generator consists of two joined spheres, each of approximately $4.5 \ \m$ in diameter and mounted on $7.5 \ \m$ tall poles, producing an average voltage of $2 \ \MV$ at its surface. This implies that mCP pair-production can occur over a large volume, since the sourced electric field dominates over Earth's magnetic field within the large viewing auditorium out to a distance of $L \gtrsim 10 \ \m$. As representative conservative values, we therefore take $E \sim 0.5 \ \MV / \m$ and $L \sim 5 \ \m$. From \Eq{qxLL2}, this implies that an mCP condensate quickly forms for $\mx \ll 10^{-8} \ \eV$ and $\qx \gtrsim 10^{-14}$, shorting-out the generator well within the time of $\sim 1 \ \s$ it takes to charge and discharge the device on a nearby conductor. The resulting upper bound on the mCP charge is shown as the line labelled ``Van de Graaff'' in \Fig{reach}. Note that this is comparable to existing astrophysical limits derived from considerations of stellar evolution~\cite{Davidson:2000hf,Vogel:2013raa,Fung:2023euv}.

\section{Terrestrial Electricity}
\label{sec:atmosphere}

\subsection{Thunderstorms}

\begin{figure}[t]
\centering
\includegraphics[width=0.5 \textwidth]{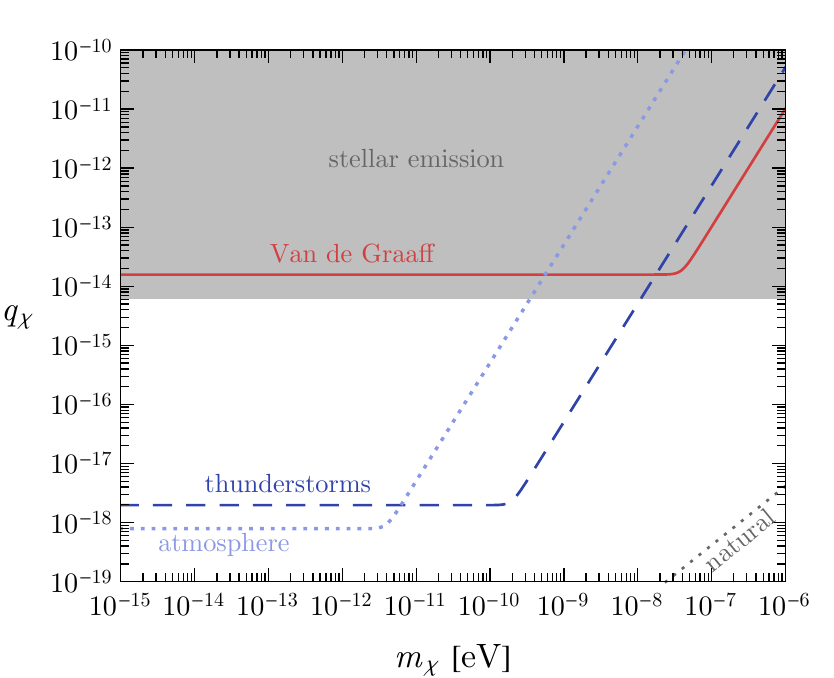}
\caption{New limits on scalar millicharged particles, as a function of charge $\qx$ and mass $\mx$, from the existence of long-ranged electric fields. These include electric fields generated by the air-insulated Van de Graaff generator at the Boston Museum of Science (solid red), thunderstorms (dashed dark blue), and the fair-weather atmosphere (dotted light blue), which would be shorted-out by the production of millicharged particles on microsecond, millisecond, and second timescales, respectively. The first two of these limits are valid for millicharged particles that directly couple to electromagnetism, or indirectly couple via an ultralight dark photon with $\order{1}$ kinetic mixing parameter, whereas the last of these limits solely applies to interactions mediated by $\order{1}$ kinetically-mixed dark photons with mass $10^{-13} \ \eV - 10^{-11} \ \eV$  (see \Secs{atmosphere}{direct} for additional details). Parameter space excluded from the consideration of stellar evolution is shown as the shaded gray region~\cite{Davidson:2000hf,Vogel:2013raa,Fung:2023euv}. Also shown as a dotted gray line is the upper bound $\qx \lesssim (4 \pi / e) \, (\mx / \Lambda)$ as imposed by technical naturalness, assuming an ultraviolet cutoff of $\Lambda = 10 \ \TeV$.}
\label{fig:reach}
\end{figure}

Strong long-ranged electric fields are also present in Earth's atmosphere. The largest of such electrical activity occurs in thunderstorms. In particular, balloon measurements show that thunderclouds maintain fields of $E \sim 10^2 \ \kV / \m$ due to internal charge separation on length scales of $L \sim 1 \ \km$, evolving on timescales of minutes~\cite{ThunderBook}. Hence, we can roughly model the electric charge configuration within a thundercloud as that of a km-scale parallel plate capacitor. Such a charge arrangement can lead to a bound condensate of mCPs, since near either charge overdensity in the thundercloud, an mCP experiences a local minimum/maximum in its electrostatic potential.  Applying the results of \Sec{production}, we find that within a millisecond, a millicharged condensate would screen such electrical configurations for $\qx \gtrsim 10^{-18}$ and $\mx \lesssim 10^{-10} \ \eV$. The resulting bound is shown as the line labelled ``thunderstorms'' in \Fig{reach}.

%

\subsection{Atmosphere}
\label{sec:fairweather}

Atmospheric electric fields also exist during fair-weather conditions, although at much smaller strengths. This consists of a downward pointing field of $E \sim 10^2 \ \V / \m$ in the $L \sim 50 \ \km$ region spanning  from the ionosphere to the crust~\cite{Feynman:1963uxa}. This voltage difference maintains equilibrium by the counteracting effects of the small global current of free ions and lightning strikes during thunderstorms, which alone would discharge or charge up the Earth on the scale of minutes, respectively. 

Note that Earth's magnetic field $B \sim 0.5 \ \text{G} \sim 10 \ \kV / \m$ near the surface is much larger than these fair-weather electric fields, which exponentially suppresses the growth of an mCP condensate directly charged under electromagnetism, as discussed in  \Sec{production}. An exception to this arises, however, if the mCP interaction is mediated by an ultralight kinetically-mixed dark photon $\Ap$, whose Compton wavelength $\sim 1 / \mAp$ is much larger than the distance $L$ between the crust and ionosphere and much shorter than the distance $r \sim 3 \times 10^3 \ \km$ to the fluid outer core from which the magnetic field is sourced. In this case, the charge separation between the crust and ionosphere generates an atmospheric dark electric field $E^\p \sim \eps \, E$, and the core sources a dark magnetic field $B^\p \sim \eps \, B \times e^{-\mAp \, r}$, where $\eps \lesssim 1$ is the dark photon kinetic mixing parameter. Thus, for dark photon masses of $10^{-13} \ \eV \ll \mAp \ll 10^{-11} \ \eV$, $E^\p$ is solely suppressed by $\eps$, whereas $B^\p \ll E^\p$ is additionally exponentially suppressed by the nonzero dark photon mass.

In this model-space, dark Schwinger production leads to an mCP condensate that instead screens $E^\p$. As a result, the condensate possesses an \emph{effective} visible charge density of $\rho_\pm \sim \eps \, E^\p / L \sim \eps^2 \, E / L$ on length scales smaller than $\sim 1 / \mAp$. Therefore, instead of screening the visible electric field by an $\order{1}$ fraction, the mCP condensate reduces it by a factor of $\sim (1 - \eps^2)$, provided that the effective visible charge of mCPs $\qx = \eps \, e^\p / e$ satisfies \Eq{qxLL2}, where $e^\p$ is the $\Ap$ gauge coupling. Note that compared to the expression provided in \Eq{gamma0}, in this case the $e$-fold rate is additionally suppressed by the relative size of the atmosphere and the radius of the Earth $\sim (50 \ \km / 10^4 \ \km) \sim 10^{-2}$, corresponding to the small probability that an mCP bound to the electric field of the Earth is within the region in which pair-production occurs.

This calculation shows that for kinetic mixing parameters close to their perturbative limit $\eps \sim 1$, a dark photon mediated mCP condensate leads to observable effects that are similar to those for a condensate that is directly charged under normal electromagnetism. The resulting bound for $\eps \sim 1$ is shown as the line labelled ``atmosphere'' in \Fig{reach}. Although  the simplest models in this parameter space are already excluded by  cosmological bounds on dark photons~\cite{Mirizzi:2009iz,Mirizzi:2009nq,Kunze:2015noa,Caputo:2020bdy,Caputo:2020rnx,Garcia:2020qrp}, in certain models such limits can be relaxed considerably by in-medium effects~\cite{Masso:2006gc,DeRocco:2020xdt,Berlin:2022hmt}.

\section{Detection of Effective mCPs}
\label{sec:direct}

The previous section showed that the millicharged condensate completely screens the electric source if the interactions are mediated directly by the Standard Model photon or indirectly by an ultralight dark photon with $\order{1}$ kinetic mixing parameter $\eps$. Instead, when $\eps \ll 1$, the sources discussed above are only screened at a fractional level of $\eps^2 \ll 1$. In this case, although large voltage sources do not exclude the existence of ultralight bosonic mCPs, they can still lead to a large number density of such dark sector particles. 

This ambient millicharge density near regions of large electrical activity leads to smaller effects that can nonetheless be detected by precision sensors of weak electromagnetic fields. As a representative example, let's first consider the final $e$-fold growth of the condensate. During this stage, there is also a burst of unbound mCPs that are repelled from the voltage source, which leads to a visible electromagnetic current of $j_\pm \sim \rho_\pm \sim \eps^2 E / L$. This current, consisting of extremely feebly interacting mCPs can traverse into shielded environments, inducing signal magnetic fields of size $B_\x \sim \eps^2 E \, R / L$, where $R$ is the size of the shielded region. Note that for $R \sim L / 10$ and $E \sim 1 \ \MV / \m$, this gives $B_\x \gtrsim 10^{-14} \ \text{T}$ for $\eps \gtrsim 10^{-5}$, which is above the threshold sensitivity of existing SQUID magnetometers~\cite{PhysRevLett.110.147002}.  

It may also be possible to directly detect the asymptotic charge density of bound mCPs using the recently proposed ``direct deflection'' setup, as studied in Refs.~\cite{Berlin:2019uco,Berlin:2021kcm}. In this case, mCP densities are detected by distorting their local flow with a region involving an oscillating electric field (note that in this case, this oscillating electric field is, in principle, distinct from the large quasistatic field responsible for the production of mCPs). This phase-space distortion leads to small oscillating electromagnetic fields, which can then be measured with a shielded resonant detector, such as an LC circuit. 

In Ref.~\cite{Berlin:2019uco}, it was found that a local dark matter density consisting of mCPs could be detected with such a setup for dark matter charges as small as $q\subdm \sim 10^{-14} \times (\mdm / \text{keV})$, where $\mdm$ is the dark matter mass. As discussed in Refs.~\cite{Berlin:2019uco,Berlin:2021kcm}, we can recast the reach of a direct deflection experiment by equating the mCP-induced Debye masses in either case,
\be
\label{eq:debye}
\frac{e \qx \, \rho_\pm}{\px} \sim \frac{(e q\subdm)^2 \, \rhodm}{(\mdm \vdm)^2}
~,
\ee
where $\rhodm \simeq 0.4 \ \GeV / \cm^3$ and $\vdm \sim 10^{-3}$ are the local dark matter mass density and velocity, respectively. Taking $\rho_\pm \sim \eps^2 \, E / L$, $p_\x \sim \sqrt{e \qx \, E}$, and $\qx \sim (e E \, L^2)^{-1}$ in \Eq{debye}, we find that this is sensitive to kinetic mixing parameters as small as $\eps \sim 10^{-10} \times (L / 10 \ \m)$. Note that this is only a parametric estimate. A more detailed study, incorporating additional effects from, e.g., the detailed velocity distribution of pair-produced mCPs or quantum pressure, is left to future work.

\section{Discussion}
\label{sec:conclusion}

The existence of extremely light millicharged particles has been extensively tested by laboratory, astrophysics, and cosmology based observations. Thus, if such particles exist in nature, their coupling to electromagnetism is constrained to be many orders of magnitude smaller than that of standard charged particles. Regardless, as shown in this work, such particles may still give rise to drastic modifications to electromagnetic observables. In particular, a charged condensate of ultralight bosonic millicharges can quickly develop within microscopic timescales, shorting out long-ranged electric fields on Earth. 

We have used this insight to place new bounds on such particles from the successful operation of large electrostatic generators in the lab and the existence of long-ranged electric sources in the atmosphere. If such particles directly couple to the photon, such bounds extend to couplings many orders of magnitude beyond that probed by astrophysics for millicharges lighter than $\sim 10^{-9} \ \eV$. 

These limits typically do not apply if the millicharges instead indirectly couple to the visible sector through an ultralight dark photon, since the effect may be suppressed by the small degree of kinetic mixing. However, even in this case, a large density of millicharges is generated, which may be directly detectable with precision electromagnetic detectors. We have discussed a few concrete examples along these lines involving magnetometers and resonant detectors. 

In this study, we have implemented order of magnitude estimates when determining the growth time and overall charge of the generated condensate. In future work, it would be interesting to perform more detailed calculations to determine the phase space and spatial profile of these millicharged condensates, incorporating the particular electric field configurations arising from the various sources discussed here. 

We also note that analogous Bose condensates may also develop in other theories, such as those incorporating new long-ranged forces coupled to baryon or lepton number. Here, we briefly outline this concept in the case of a new long-ranged $U(1)_{B-L}$ force, postponing a more detailed investigation to future work. Static tests of the equivalence principle bound the corresponding coupling to Standard Model currents to be $g_{_\text{SM}} \lesssim 10^{-5} \times (m_p / M_\text{Pl})$, where $m_p$ is the proton mass and $M_\text{Pl}$ is the Planck mass~\cite{Wagner:2012ui,Graham:2015ifn}. For couplings saturating these bounds and if the mass of the $B-L$ force-carrier is smaller than an inverse Earth radius $R_\oplus^{-1} \sim 10^{-14} \ \eV$, the entire Earth contributes to a $U(1)_{B-L}$ electric field of strength $E_{B-L} \sim 0.1 \ \MV / \m$. Furthermore, due to the small average velocity of terrestrial atoms, the $U(1)_{B-L}$ magnetic field sourced by the Earth is suppressed in comparison, $B_{B-L} \ll E_{B-L}$.  Thus, if there exist scalar particles $\x$ directly charged under this new gauge symmetry with, e.g., mass $m_\x \lesssim R_\oplus^{-1} \sim 10^{-14} \ \eV$ and coupling $g_\x \gtrsim (E_{B-L} \, R_\oplus^2)^{-1} \sim 10^{-26}$, then a $U(1)_{B-L}$ condensate quickly develops at an $e$-fold rate of $\Gamma_\x \sim 100 \ \Hz$, shorting out Earth's field to a subcritical value $E_{B-L} \lesssim \max ( B_{B-L} \, , \, m_\x^2 / g_\x)$. This potentially has important implications for laboratory tests of new forces, which will be investigated in future work.


\section*{Acknowledgements}
We would like to thank Masha Baryakhtar, Paddy Fox, Anson Hook, and Davide Racco for useful conversations, as well as Cristina Mondino, Katelin Schutz, and Ken Van Tilburg for valuable collaboration on ongoing related work. Fermilab is operated by the Fermi Research Alliance, LLC under Contract DE-AC02-07CH11359 with the U.S. Department of Energy. This material is based upon work supported by the U.S. Department of Energy, Office of Science, National Quantum Information Science Research Centers, Superconducting Quantum Materials and Systems Center (SQMS) under contract number DE-AC02-07CH11359. This work was completed in part at the Perimeter Institute. Research at Perimeter Institute is supported in part by the Government of Canada through the Department of Innovation, Science and Economic Development Canada and by the Province of Ontario through the Ministry of Colleges and Universities. Y.-Y. L is supported by the NSF of China through Grant No. 12305107, 12247103.

\bibliography{main}


\end{document}